\documentclass[rnote]{aa}
\usepackage[varg]{txfonts}
\usepackage{color}

\bibpunct{(}{)}{;}{a}{}{,} 

\begin{document}


\title{A hot bubble at the centre of M81}


\author{T.V. Ricci
\and J.E. Steiner 
\and L. Giansante }

\institute{Instituto de Astronomia, Geof\'isica e Ci\^encias Atmosf\'ericas, Universidade de S\~ao Paulo, 05508-900, S\~ao Paulo, SP, Brasil \\ \email{tvricci@iag.usp.br, joao.steiner@iag.usp.br, louise.martins@usp.br}}

\date{Received <date> /
Accepted <date>}

\abstract{\textit{Context.} Messier 81 has the nearest active nucleus with broad H$\alpha$ emission. A detailed study of this galaxy's centre is important for understanding the innermost structure of the AGN phenomenon.

\textit{Aims.} Our goal is to seek previously undetected structures using additional techniques to reanalyse a data cube obtained with the GMOS-IFU installed on the Gemini North telescope \citep{2011MNRAS.413..149S}.

\textit{Method.} We analysed the data cube using techniques of noise reduction, spatial deconvolution, starlight subtraction, PCA tomography, and comparison with HST images.

\textit{Results.} We identified a hot bubble with T $>$ 43500 K that is associated with strong emission of [N II]$\lambda$5755\AA\ and a high [O I]$\lambda$6300/H$\alpha$ ratio; the bubble displays a bluish continuum, surrounded by a thin shell of H$\alpha$ + [N II] emission. We also reinterpret the outflow found by \citet{2011MNRAS.413..149S} showing that the blueshifted cone nearly coincides with the radio jet, as expected. 

\textit{Conclusions.} We interpret the hot bubble as having been caused by post starburst events that left one or more clusters of young stars,  similar to the ones found at the centre of the Milky Way, such as the Arches and the IRS 16 clusters. Shocked structures from combined young stellar winds or supernova remnants are probably the cause of this hot gas and the low ionization emission. 
}
\maketitle

\section{Introduction}

Messier 81 (NGC3031) is an SA(s)ab galaxy, known to host a low ionization nuclear emission region (LINER) \citep{1980A&A....87..152H} associated with a type 1 active galactic nucleus (AGN) \citep{1981ApJ...245..845P}. At a distance of 3.5 Mpc, this makes it the closest case of a known type 1 AGN. Radio emission in the form of a compact core, a nuclear jet \citep{2000ApJ...532..895B,2011A&A...533A.111M}, and nuclear X-ray emission \citep{2007ApJ...669..830Y} are associated with this AGN.

Although photoionization by an AGN is one of the possible causes of excitation for LINERs \citep{1983ApJ...264..105F,1983ApJ...269L..37H}, shocks \citep{1980A&A....87..152H} and hot old stellar populations \citep{1994A&A...292...13B} have also been proposed as potential sources. Given its distance, this is an important object for studying emission line properties with high spatial resolution.

In this research note, we reanalyse a data cube of M81 \citep{2011MNRAS.413..149S} obtained with the integral field unit (IFU) of the Gemini Multi-Object Spectrograph (GMOS) installed on the Gemini North telescope. We report the detection of hot (T$>$40000 K) gas, located $\sim$ 0.3 to 0.8 arcsec (5 to 14 pc) southward from the nucleus of this galaxy. We also reinterpret the spatial orientation of the nuclear gaseous disc and outflow reported by \citet{2011MNRAS.413..149S}\footnote{Allan Schnorr-M\"{u}ller was contacted about this new result in M81 and also about the correction to the spatial orientation of the data cube.}.

\section{Observations and data reduction} \label{observation_data_reduction_section}

M81 was observed with the Gemini North telescope on 2006 December 31 (programme GN-2006B-Q-94). We used the GMOS-IFU in two-slit mode. Three exposures of 530s covered the central region of M81, with a field of view (FOV) of 7 x 5 arcsec$^2$. The R400 grating was used, resulting in an observed wavelength range of 5600-7000\AA\ and spectral resolution R $\sim$ 1800. The raw data were reduced with the standard Gemini package under the {\sc IRAF} environment. Bias, flat-fields, calibration lamps, and spectrophotometric standards were obtained for the overall correction and calibration of the data. Cosmic rays were removed with the {\sc LACOS} algorithm \citep{2001PASP..113.1420V}. Finally, three flux-calibrated data cubes were created with a spatial sampling of 0.05 arcsec.

In addition to the basic reduction steps, we also adopted additional procedures to improve the quality of the data cubes \citep{2014MNRAS.440.2419R,2014MNRAS.438.2597M}. First, we corrected each data cube for the effects of differential atmospheric refraction (DAR) using an algorithm developed by us. Secondly, we calculated the median of the three data cubes to avoid bad pixels and cosmic rays that had not been properly removed by {\sc LACOS}. We then removed high-frequency noise from the spatial dimensions by convolving each image of the median data cube with a Butterworth low-pass filter \citep{2008gonzaleswoods}, with a spatial-frequency cut of 0.2F$_{NY}$, where F$_{NY}$ = 0.5 spaxels$^{-1}$ is the Nyquist frequency and n = 2 (see \citealt{2014MNRAS.440.2419R} for more details on the Butterworth filtering process). Instrumental fingerprints were removed with principal component analysis (PCA) tomography \citep{2009MNRAS.395...64S,2014MNRAS.440.2419R,2014MNRAS.438.2597M}. Finally, we deconvolved each image of the median data cube using a Richardson-Lucy algorithm \citep{1972JOSA...62...55R,1974AJ.....79..745L}, with ten iterations, assuming a Moffat point spread function (PSF) with full width at half maximum (FWHM) of 1.0 arcsec and $\beta$ = 2.9. The deconvolved PSF is described by a Gaussian function with FWHM = 0.89 arcsec, measured in the image of the broad component of the H$\alpha$ emission line.

\section{PCA tomography}

PCA tomography \citep{2009MNRAS.395...64S} consists of applying PCA to data cubes with the aim of extracting useful information and removing instrumental noise \citep[see e.g. ][]{2011ApJ...734L..10R,2041-8205-765-2-L40,2014MNRAS.440.2419R,2014MNRAS.438.2597M}. This method allows the separation of information originally expressed in a system of correlated coordinates into a system of uncorrelated coordinates (eigenvectors or eigenspectra). Tomograms are projections of eigenspectra on the data cubes and reveal where the correlations between the wavelengths occur on the spatial dimension of the data cubes.

\begin{figure}
        \resizebox{\hsize}{!}{\includegraphics{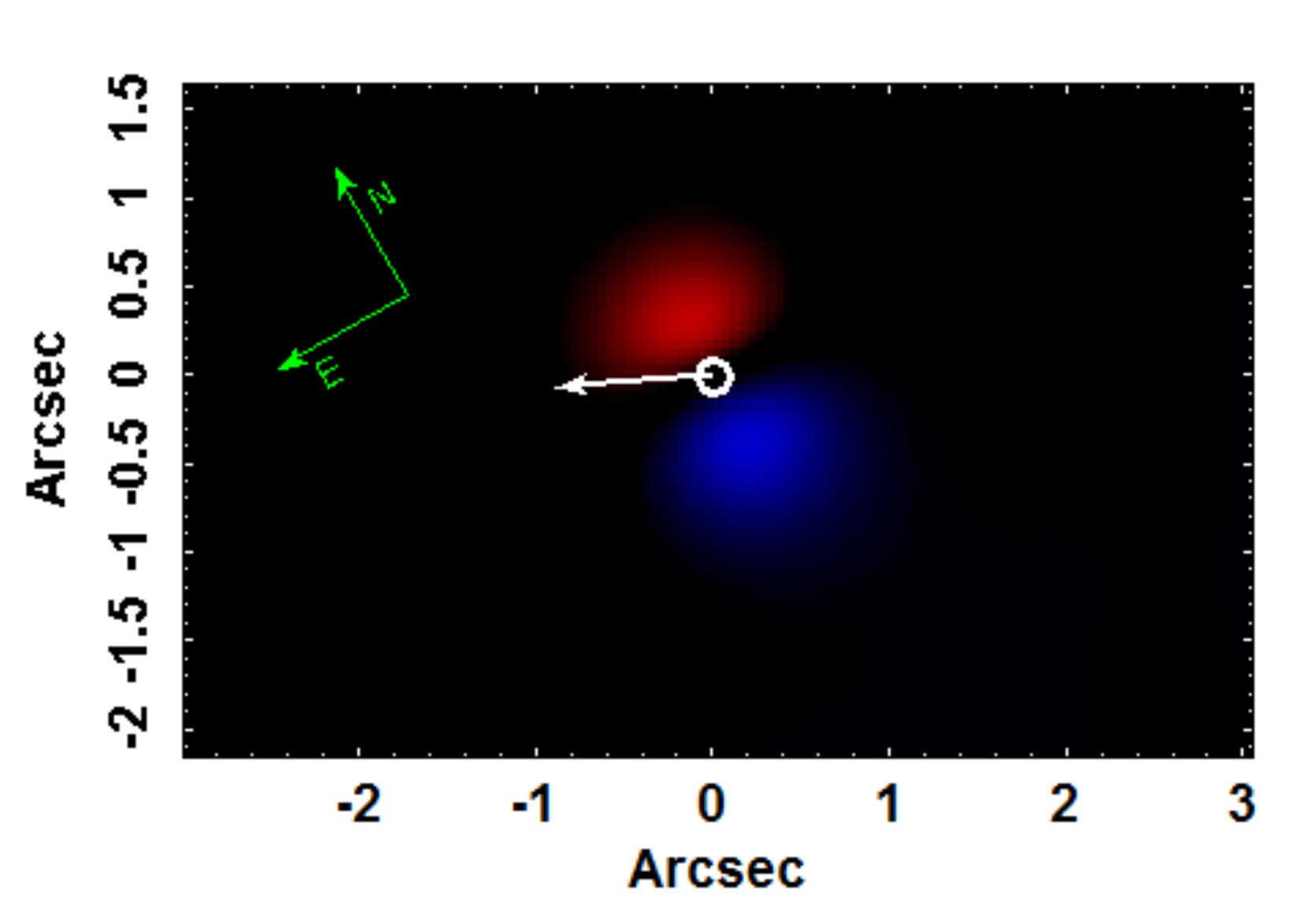}}
        \resizebox{\hsize}{!}{\includegraphics{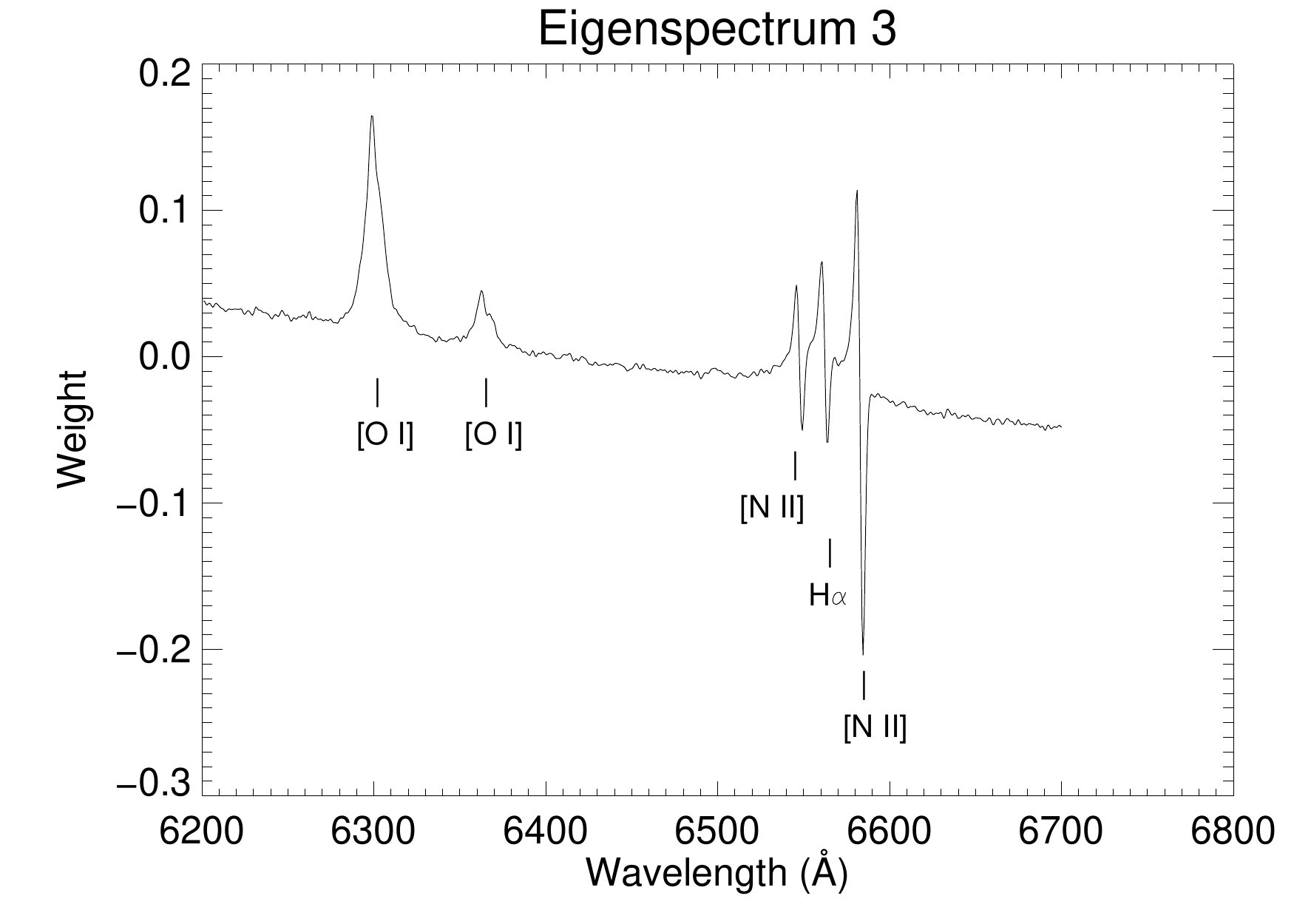}}
        \caption{Tomogram and eigenspectrum 3 of the PCA tomography applied to the 6200-6700 \AA\ spectral range. The anti-correlation between the red and the blue wings of the [N II] and H$\alpha$ emission lines associated with a bipolar structure seen in the tomogram is a typical signature of gas kinematics. In the tomogram, the structure in blue (red) corresponds to the blue (red) wings of the [N II] and H$\alpha$ emission lines. In this case, we interpreted this kinematics as a gas disc with P.A. = -3$^{o} \pm$1$^o$. The white circle marks the position of the AGN as given by the image of the broad component of H$\alpha$, and the white arrow corresponds to the direction of a resolved nuclear radio jet with P.A. = 65$^o$ \citep{2011A&A...533A.111M}.}
        \label{tom_eig_gas_disc}
\end{figure}

\citet{2011MNRAS.413..149S} applied PCA tomography to the data cube of M81 in the 6200-6700 \AA\ spectral range. They argued that the bipolar structure seen in the tomogram related to the third eigenspectra is associated with a gas disc. We agree with their interpretation; however, because of an error in the spatial orientation of the FOV (the x-axis of these authors corresponds to -x in our figures), the position angle (P.A.) of the gas disc is now -3$^{o} \pm$1$^o$, measured as the angle of the line that connects the positions of minimum and maximum weights in the tomogram. We show the tomogram and the eigenspectrum related to the gas disc in Fig. \ref{tom_eig_gas_disc}. It is worth noticing that, in addition to the disc-like kinematics associated with the H$\alpha$+[N II] lines, one can also see a strong correlation between [O I] lines and the blueshifted side of the disc to the south of the nucleus. Moreover, a significant bluish continuum is also related to this feature. Such correlations have not been seen in other disc-like kinematics of early-type galaxies \citep{2014MNRAS.440.2419R}.

\begin{figure}
        \resizebox{\hsize}{!}{\includegraphics{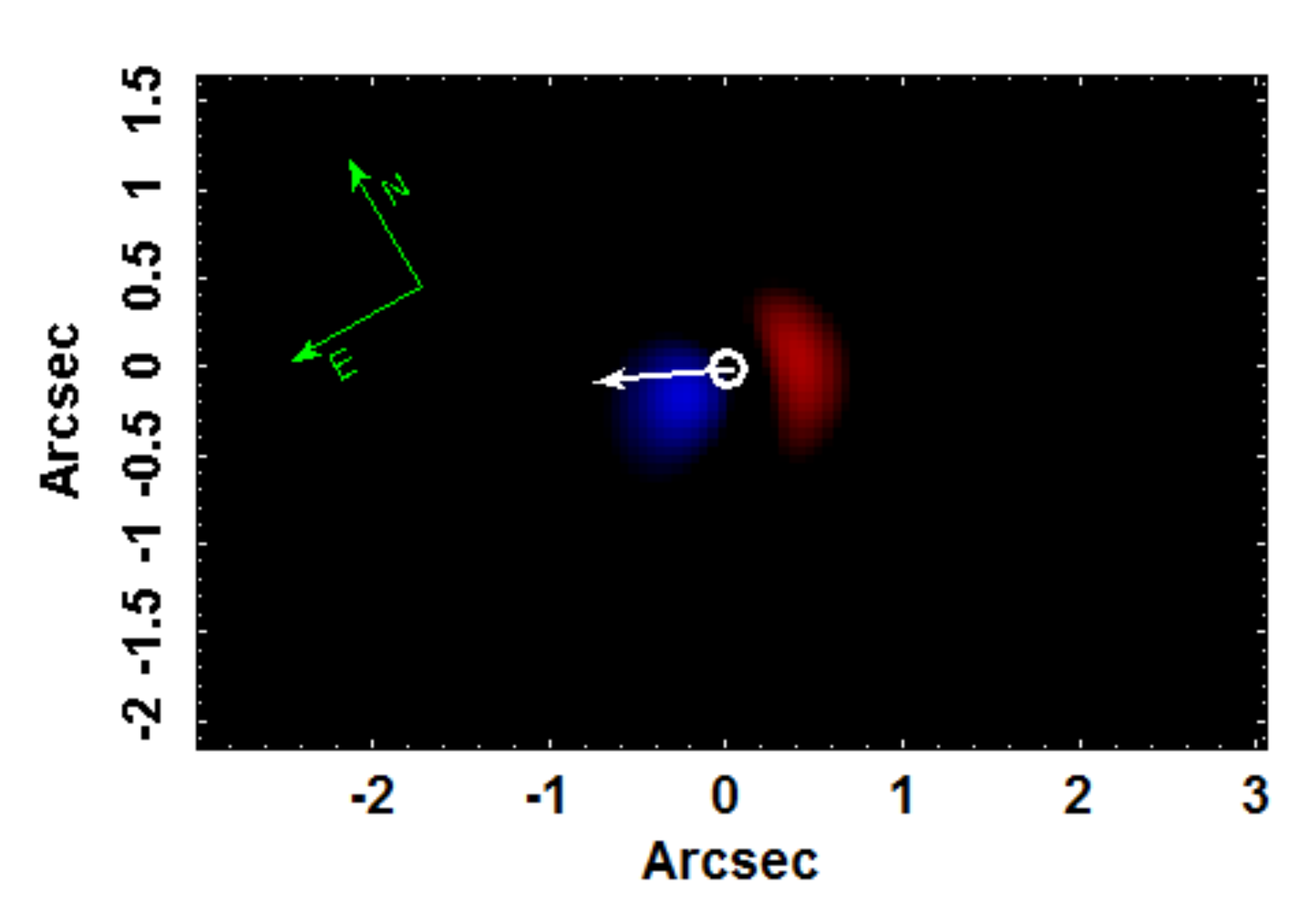}}
        \resizebox{\hsize}{!}{\includegraphics{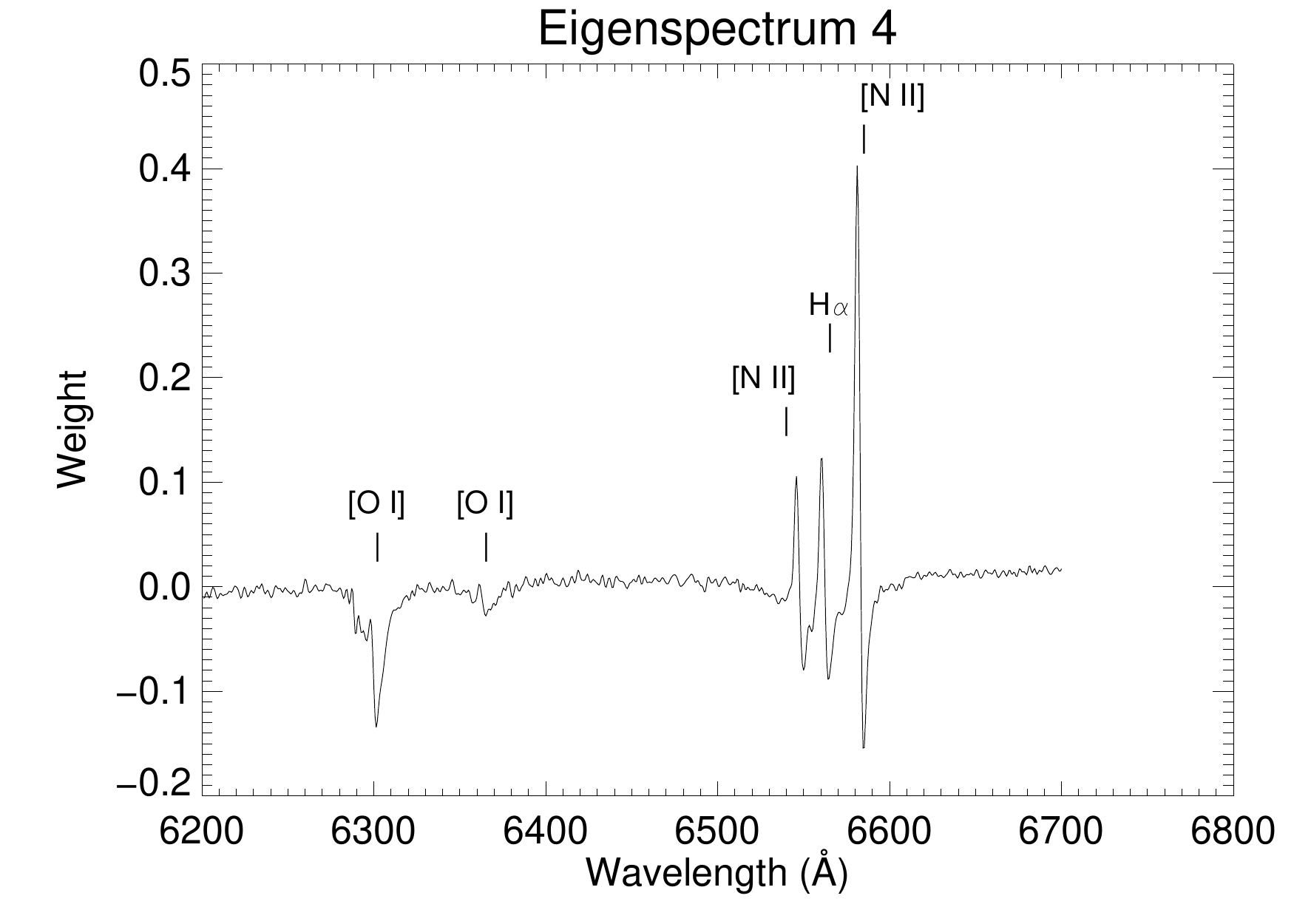}}
        \caption{Tomogram and eigenspectrum 4 of the PCA tomography applied to the 6200-6700 \AA\ spectral range. In this case, the gas kinematics is probably associated with outflows. We note that the bipolar structure is almost perpendicular to the gas disc (P.A. = -103$^o \pm$ 4$^o$). In addition, the nuclear radio jet is in the same direction (P.A. = 65$^o$) as the component of the outflow that is in blueshift relative to the nucleus of M81.}
        \label{tom_eig_outflow}
\end{figure}

The tomogram and the eigenspectrum 4, shown in Fig. \ref{tom_eig_outflow}, also suggest the motion of the gas, but the direction of this kinematic feature is almost perpendicular to the gas disc (P.A. = -103$^o \pm$ 4$^o$). This is probably related to an outflow, since a nuclear radio jet is pointing to the same direction as the component that is in blueshift relative to the nucleus \citep[P.A. = 65$^o$:][]{2011A&A...533A.111M}. \citet{2011MNRAS.413..149S} also proposed that the fourth eigenspectrum is related to an outflow; however, in their study, because of the incorrect spatial orientation of the FOV, the nuclear radio jet pointed to a structure that was in redshift relative to the nucleus of M81. 

\begin{figure}
        \resizebox{\hsize}{!}{\includegraphics{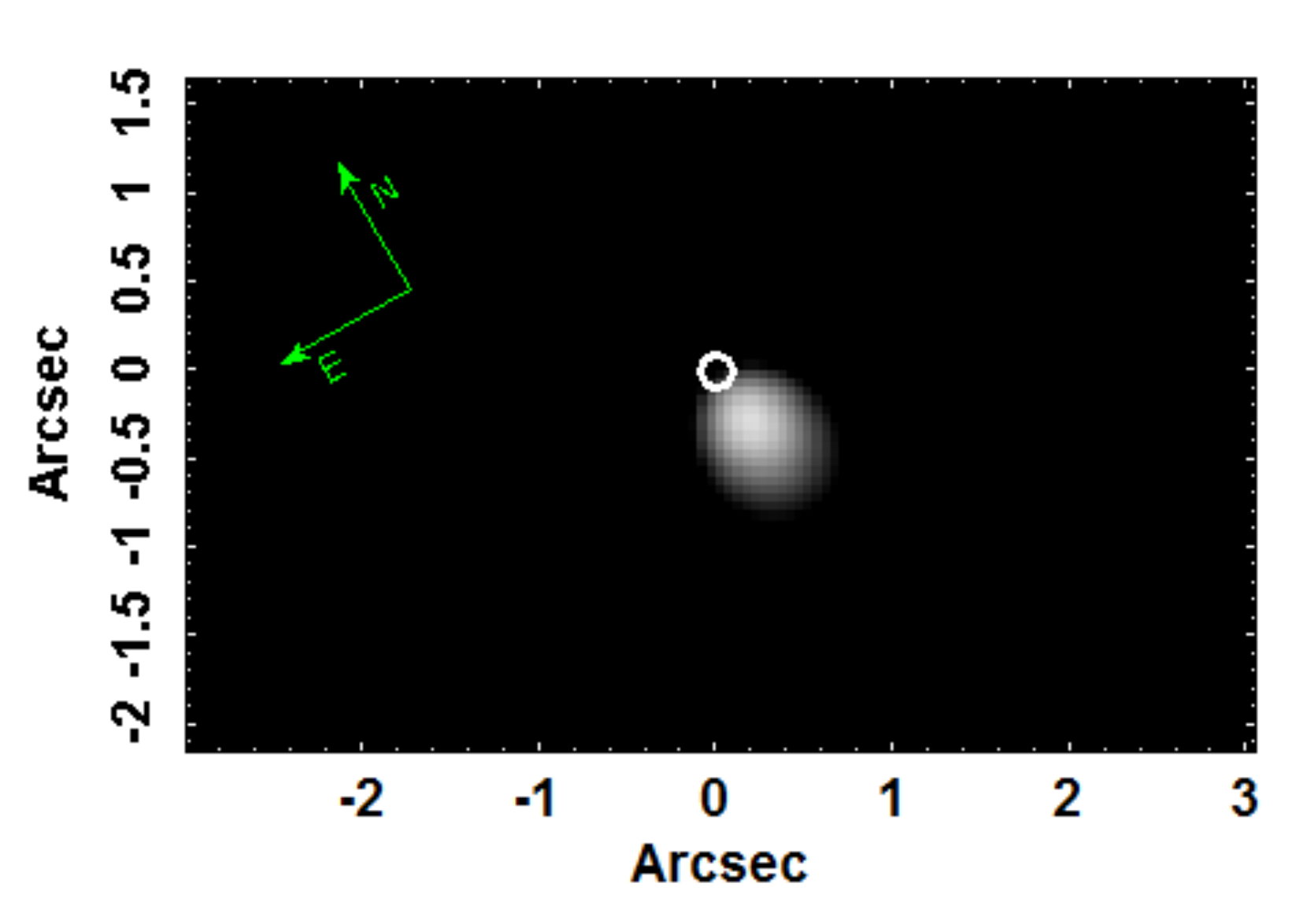}}
        \resizebox{\hsize}{!}{\includegraphics{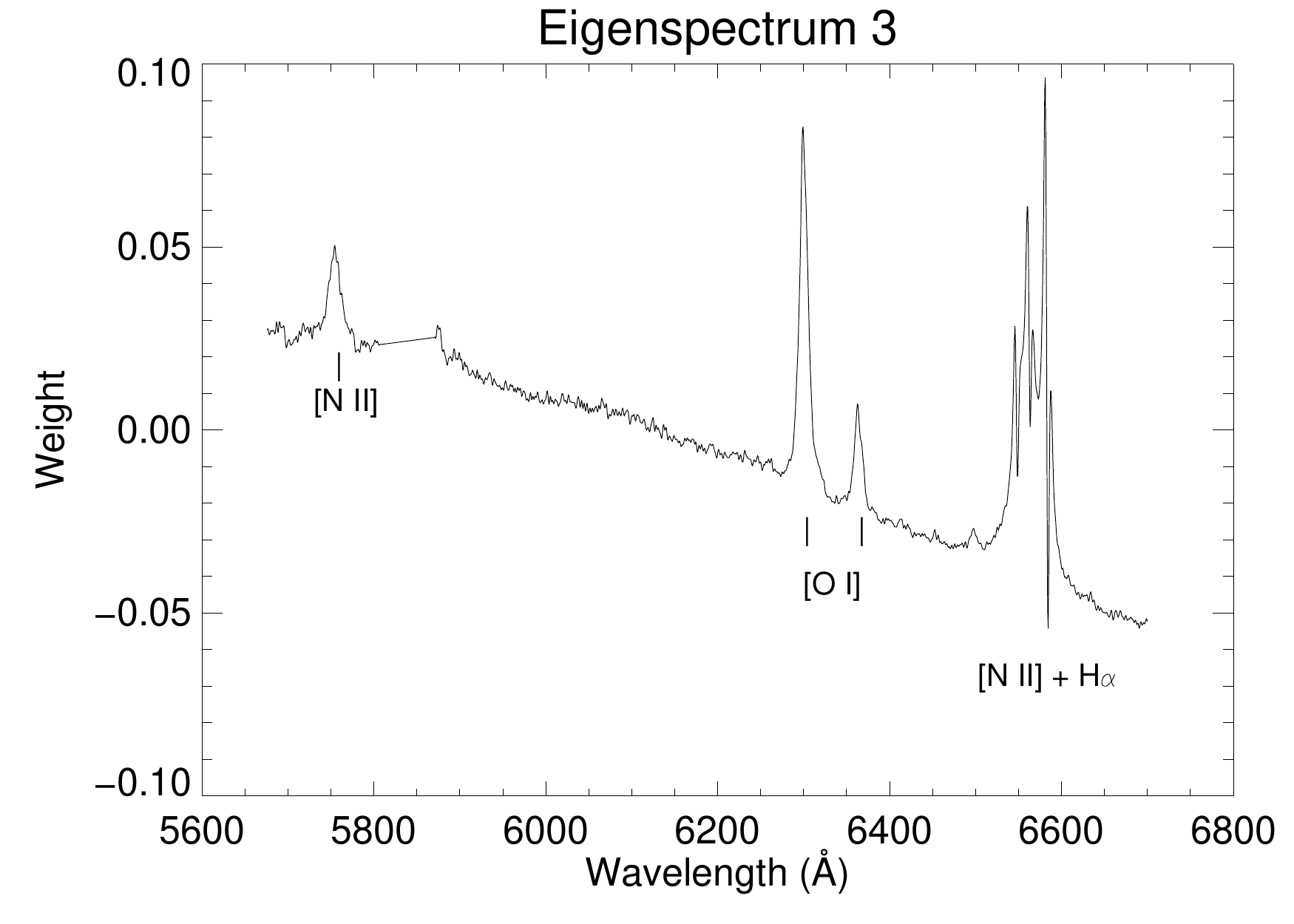}}
        \caption{Tomogram and eigenspectrum 3 of the PCA tomography applied to the 5675-6700 \AA\ spectral range. This range includes the [N II]$\lambda$5755 emission line, which is sensitive to the high temperatures of the ionized gas. The eigenvector reveals a correlation between all emission lines and the blue component of the continuum. }
        \label{tom_eig_hotregion}
\end{figure}

We also applied PCA tomography to the 5675-6700 \AA\ spectral range. The main difference here is that we included in the analysis the [N II]$\lambda$5755 emission line, which is sensitive to the high temperatures of the ionized gas. Again, the first and second eigenspectra are related to the galactic bulge and to the AGN, respectively. However, the third eigenspectrum, shown in Fig. \ref{tom_eig_hotregion}, reveals a correlation between the [N II]$\lambda$5755, [O I]$\lambda\lambda$6300,6363, and H$\alpha$+[N II]$\lambda\lambda$6548,6583 emission lines. These lines are also correlated with the blue region of the continuum. The tomogram related to this eigenspectrum, also shown in Fig. \ref{tom_eig_hotregion}, reveals a region associated with these emission lines south of the nucleus, again coinciding with the blueshifted side of the gaseous disc. The interpretation is that this region is composed of very hot gas and that the continuum is bluer than the overall stellar component along the FOV, which indicates the presence of young stellar populations. 

\section{Emission line properties}

In order to study emission lines emerging from the hot gas detected with PCA tomography, we subtracted the stellar components of each spectrum of the data cube by means of the spectral synthesis technique. We used the software {\sc STARLIGHT} \citep{2005MNRAS.358..363C} with the stellar population base proposed by \citet{2009MNRAS.398L..44W}, which contains 120 stellar populations with ages between 3 and 12 Gyr; abundances of [Fe/H] = -0.5, -0.25, 0.0, and 0.2; and [$\alpha$/Fe] = 0.0, 0.2, and 0.4. The result of subtracting the solutions of the spectral synthesis from the spectra of the data cube for each spaxel is a gas cube, i.e. a data cube containing only the gas component. Since we are interested in analysing only the narrow component of the emission lines, we removed the broad component of the H$\alpha$ line from the gas cube of M81 in a way similar to that of \citet{2014MNRAS.440.2442R}.

\begin{figure}
        \resizebox{\hsize}{!}{\includegraphics{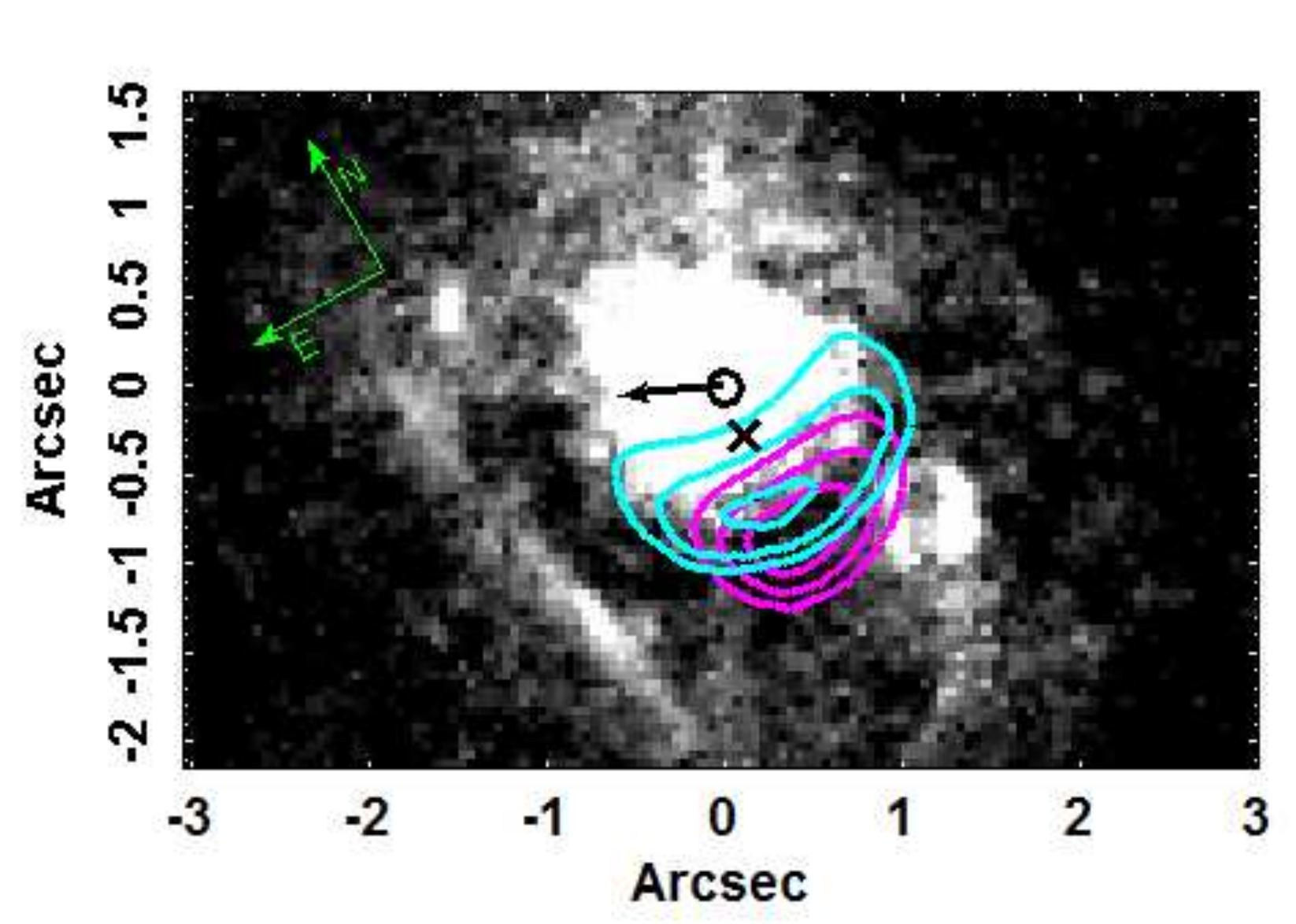}}
        \caption{(H$\alpha$+[N II])/I ratio using images from the ACS aboard the HST. We note that the peak of the emission coincides with the position of the nucleus. The magenta contours are curves of isotemperatures. The outer contour has T$_e$ $\sim$ 43500K and the temperature increases inward. The cyan contours are associated with the [O I]/H$\alpha$ ratio, with values of 0.6, 0.7, and 0.8 from the outside to the inside. The $\times$ sign marks the maximum [N II]$\lambda$5755 emission. A very hot region associated with high values of [O I]/H$\alpha$ is probably related to shock-heated regions since photoionization models cannot produce such a high temperature. The resolved nuclear radio jet has a P.A. of 65$^o$.}
        \label{hst_temperature_oi_ha}
\end{figure}

We measured the [O I]$\lambda$6300/H$\alpha$ and [N II]$\lambda$5755/[N II]$\lambda\lambda$6548+6584 line ratios along the FOV. To do so, we adjusted the profile of each line with a Gauss-Newton algorithm to fit non-linear functions. For each spaxel, we assumed that the [O I], [N II], and H$\alpha$ lines have the same velocity and FWHM. In Fig. \ref{hst_temperature_oi_ha} we show the [O I]/H$\alpha$ line ratio map. We also show the gas temperature derived from the [N II]$\lambda$5755/[N II]$\lambda\lambda$6548+6584 \citep{2006agna.book.....O}, assuming a low gas density regime. The outermost isothermal curve (red line in Fig. \ref{hst_temperature_oi_ha}) corresponds to T$_e$ = 43500 K. The maximum emission of [N II]5755 is located at 0.29 $\pm$ 0.07 arcsec south of the nucleus, while the [O I]/H$\alpha$ ratio is at 0.7 arcsec, where it reaches $\gtrsim$ 0.6, and the maximum of the temperature seems to be located at 0.8 arcsec. We conclude that the bubble comprises a region from 5 to 14 pc from the nucleus of M81. Given the limitations of our resolution it is not possible to assert its real dimension, although the HST H$\alpha$ emission (Fig. \ref{hst_temperature_oi_ha}) suggests that the bubble could extend as far as 2 arcsec (35 pc). It would be highly desirable to have observations with higher spatial resolution, for example with adaptive optics of the [Fe II]$\lambda\lambda$12567, 16435 lines. 

The [N II]$\lambda$5755 line has a width of 388 $\pm$ 9 km s$^{-1}$; we do not have an indication of the gas density because the [S II] doublet fell in the spectral gap of the CCD detectors. Therefore, no mass estimate can be made.  

In Fig. \ref{hst_temperature_oi_ha} we also show the ratio between the archival images of the narrow H$\alpha$+[N II]$\lambda\lambda$6548+6584 filter (namely F658N) and of the I band obtained with the Wide Field Camera (WFC) of the Advanced Camera for Surveys (ACS) on board the Hubble Space Telescope (HST). This ratio shows the strong emission of the H$\alpha$+[N II]$\lambda\lambda$6548+6584 lines the peak of which is taken to be the centre of the FOV, as it presumably indicates the position of the AGN. For the GMOS frame, we assume the centre to be derived from the broad H$\alpha$ red wing. With this assumption we can overlay images of both the HST and Gemini. In addition to the AGN, a faint shell of emission nearly encircling the region of hot gas is also seen in the HST image. The maximum temperature is located 0.8 arcsec (14 pc) south of the nucleus.

\section{Discussion and conclusions}

In this work, we reanalyse the data cube of M81 studied by \citet{2011MNRAS.413..149S}. The authors found that this galaxy has a gaseous disc as well as an outflow associated with the AGN. We found that their x-axis was inverted from what it should have been and, therefore, that the new P.A. of these gaseous features are quite distinct. In particular, the radio jet observed by \citet{2000ApJ...532..895B} and \citet{2011A&A...533A.111M} seems to be associated with the redshifted cone/outflow structure (ionization cone) detected by \citet{2011MNRAS.413..149S}; for this to make sense, one has to assume that a hypothetical counter-jet does not exist. 

With our new orientation of the FOV, the blueshifted cone/outflow is co-aligned with the radio jet, for which we assume a new P.A., given by \citet{2011A&A...533A.111M} and shown in Fig. \ref{tom_eig_outflow}. The asymmetry of an one-sided radio jet is now naturally explained by the Doppler boosting effect.

When we analysed the data cube with the [N II]$\lambda$5755\AA\ line (not included in the \citealt{2011MNRAS.413..149S} study), we detected an unresolved hot spot located 0.8 arcsec south of the nucleus. We estimate a temperature T$_e$ $>$ 43500 K for the gas in this region. The line ratio [O I]/H$\alpha$ = 0.8 is exceptionally high, even for LINERs \citep{1997ApJS..112..315H}; it is also higher than on the nucleus of M81, where the ratio is [O I]/H$\alpha$ = 0.6. High temperatures, associated with high [O I]/H$\alpha$, are typical signatures of mechanical heating such as occurs in shock waves produced, for example in supernova remnants (SNR) or shocked stellar winds \citep{1996ApJS..102..161D,2006agna.book.....O,2008ApJS..178...20A}. An alternative would be a shock wave produced by a superwind induced by the AGN. This interpretation, however, seems to be at odds with the jet orientation. The bubble is located perpendicular to the direction of the jet; this suggests that the bubble is not associated with outflows from the AGN.

The structure detected in M81 may be similar to (but on a larger scale)  the structure known as Arches Cluster \citep{2002ApJ...570..665Y}, located about 30 pc from the centre of the Milky Way, or to the young stellar cluster associated with IRS 16, even closer to the centre of our Galaxy (see \citealt{2010RvMP...82.3121G} for a review). In particular, the Arches Cluster shows a structure of cold gas surrounding it. At the same time, hot intracluster gas was detected by its X-ray emission and was interpreted as a combined shock of stellar winds from young (1 - 3 Myr) stars \citep{2002ApJ...570..665Y}. In M81, as in the Arches Cluster, the hot bubble is also encircled by a filament of cooler gas emission (Fig. \ref{hst_temperature_oi_ha}). Here, young stellar populations revealed by PCA tomography may be associated with a more massive cluster or, perhaps, a combination of many clusters. For example,  \citet{2014ApJ...784...79D} detected four young stellar clusters encircling the AGN of the galaxy  NGC 2110. Using IFU data with adaptive optics in the IR, they associated [Fe II] emissions with shocks caused by winds and/or SNR emerging from these clusters. In fact, they showed that the peak of the [Fe II] emission in the circumnuclear region of NGC 2110 is not collocated with the stellar clusters, but lies between them, which indicates that the shocks occur in the intracluster medium. In addition to the shocked gas from young stellar winds, shock waves from SNR of somewhat older (age $<$ 70 Myr) stellar populations may also be at play in M81. 

Since M81 is a nearby galaxy, it is possible to discuss the importance of shocks in LINERs even when an AGN is clearly detected. Because of the proximity of M81, we were able to resolve the AGN and a region dominated by shocks. If the same scenario applies to relatively distant galaxies, both regions would be unresolved and probably classified as pure LINER-like AGNs, i.e. with no evidence of shocks. The case of M81 suggests that shocks may have an important role even in LINERs that are clearly ionized by other sources, for instance AGNs or post-AGB stars.

\begin{acknowledgements} 

This paper was based on observations obtained at the Gemini Observatory acquired through the Gemini Science Archive and processed using the Gemini IRAF package, which is operated by the Association of Universities for Research in Astronomy, Inc., under a cooperative agreement with the NSF on behalf of the Gemini partnership: the National Science Foundation (United States), the National Research Council (Canada), CONICYT (Chile), the Australian Research Council (Australia), Minist\'{e}rio da Ci\^{e}ncia, Tecnologia e Inova\c{c}\~{a}o (Brazil) and Ministerio de Ciencia, Tecnolog\'{i}a e Innovaci\'{o}n Productiva (Argentina). This work is also based on observations made with the NASA/ESA Hubble Space Telescope, and obtained from the Hubble Legacy Archive, which is a collaboration between the Space Telescope Science Institute (STScI/NASA), the Space Telescope European Coordinating Facility (ST-ECF/ESA) and the Canadian Astronomy Data Centre (CADC/NRC/CSA). T.V. Ricci and L. Giansante also acknowledge FAPESP for financial support under grants 2012/21350-7 (T.V.R.) and 2011/20224-5 (L.G.)

\end{acknowledgements}

\bibliographystyle{aa}
\bibliography{bibliografia}

\begin{thebibliography}{28}
\expandafter\ifx\csname natexlab\endcsname\relax\def\natexlab#1{#1}\fi

\bibitem[{{Allen} {et~al.}(2008){Allen}, {Groves}, {Dopita}, {Sutherland}, \&
  {Kewley}}]{2008ApJS..178...20A}
{Allen}, M.~G., {Groves}, B.~A., {Dopita}, M.~A., {Sutherland}, R.~S., \&
  {Kewley}, L.~J. 2008, \apjs, 178, 20

\bibitem[{{Bietenholz} {et~al.}(2000){Bietenholz}, {Bartel}, \&
  {Rupen}}]{2000ApJ...532..895B}
{Bietenholz}, M.~F., {Bartel}, N., \& {Rupen}, M.~P. 2000, \apj, 532, 895

\bibitem[{{Binette} {et~al.}(1994){Binette}, {Magris}, {Stasi{\'n}ska}, \&
  {Bruzual}}]{1994A&A...292...13B}
{Binette}, L., {Magris}, C.~G., {Stasi{\'n}ska}, G., \& {Bruzual}, A.~G. 1994,
  \aap, 292, 13

\bibitem[{{Cid Fernandes} {et~al.}(2005){Cid Fernandes}, {Mateus}, {Sodr{\'e}},
  {Stasi{\'n}ska}, \& {Gomes}}]{2005MNRAS.358..363C}
{Cid Fernandes}, R., {Mateus}, A., {Sodr{\'e}}, L., {Stasi{\'n}ska}, G., \&
  {Gomes}, J.~M. 2005, \mnras, 358, 363

\bibitem[{{Dopita} \& {Sutherland}(1996)}]{1996ApJS..102..161D}
{Dopita}, M.~A. \& {Sutherland}, R.~S. 1996, \apjs, 102, 161

\bibitem[{{Durr{\'e}} \& {Mould}(2014)}]{2014ApJ...784...79D}
{Durr{\'e}}, M. \& {Mould}, J. 2014, \apj, 784, 79

\bibitem[{{Ferland} \& {Netzer}(1983)}]{1983ApJ...264..105F}
{Ferland}, G.~J. \& {Netzer}, H. 1983, \apj, 264, 105

\bibitem[{{Genzel} {et~al.}(2010){Genzel}, {Eisenhauer}, \&
  {Gillessen}}]{2010RvMP...82.3121G}
{Genzel}, R., {Eisenhauer}, F., \& {Gillessen}, S. 2010, Reviews of Modern
  Physics, 82, 3121

\bibitem[{{Gonzales} \& {Woods}(2008)}]{2008gonzaleswoods}
{Gonzales}, R. \& {Woods}, R. 2008, {Digital Image Processing}

\bibitem[{{Halpern} \& {Steiner}(1983)}]{1983ApJ...269L..37H}
{Halpern}, J.~P. \& {Steiner}, J.~E. 1983, \apjl, 269, L37

\bibitem[{{Heckman}(1980)}]{1980A&A....87..152H}
{Heckman}, T.~M. 1980, \aap, 87, 152

\bibitem[{{Ho} {et~al.}(1997){Ho}, {Filippenko}, \&
  {Sargent}}]{1997ApJS..112..315H}
{Ho}, L.~C., {Filippenko}, A.~V., \& {Sargent}, W.~L.~W. 1997, \apjs, 112, 315

\bibitem[{{Lucy}(1974)}]{1974AJ.....79..745L}
{Lucy}, L.~B. 1974, \aj, 79, 745

\bibitem[{{Mart{\'{\i}}-Vidal} {et~al.}(2011){Mart{\'{\i}}-Vidal}, {Marcaide},
  {Alberdi}, {P{\'e}rez-Torres}, {Ros}, \& {Guirado}}]{2011A&A...533A.111M}
{Mart{\'{\i}}-Vidal}, I., {Marcaide}, J.~M., {Alberdi}, A., {et~al.} 2011,
  \aap, 533, A111

\bibitem[{Menezes {et~al.}(2013)Menezes, Steiner, \&
  Ricci}]{2041-8205-765-2-L40}
Menezes, R.~B., Steiner, J.~E., \& Ricci, T.~V. 2013, \apjl, 765, L40

\bibitem[{{Menezes} {et~al.}(2014){Menezes}, {Steiner}, \&
  {Ricci}}]{2014MNRAS.438.2597M}
{Menezes}, R.~B., {Steiner}, J.~E., \& {Ricci}, T.~V. 2014, \mnras, 438, 2597

\bibitem[{{Osterbrock} \& {Ferland}(2006)}]{2006agna.book.....O}
{Osterbrock}, D.~E. \& {Ferland}, G.~J. 2006, {Astrophysics of gaseous nebulae
  and active galactic nuclei} (2nd.~ed.~by D.E.~Osterbrock and
  G.J.~Ferland.~Sausalito, CA: University Science Books, 2006)

\bibitem[{{Peimbert} \& {Torres-Peimbert}(1981)}]{1981ApJ...245..845P}
{Peimbert}, M. \& {Torres-Peimbert}, S. 1981, \apj, 245, 845

\bibitem[{{Ricci} {et~al.}(2011){Ricci}, {Steiner}, \&
  {Menezes}}]{2011ApJ...734L..10R}
{Ricci}, T.~V., {Steiner}, J.~E., \& {Menezes}, R.~B. 2011, \apjl, 734, L10

\bibitem[{{Ricci} {et~al.}(2014{\natexlab{a}}){Ricci}, {Steiner}, \&
  {Menezes}}]{2014MNRAS.440.2442R}
{Ricci}, T.~V., {Steiner}, J.~E., \& {Menezes}, R.~B. 2014{\natexlab{a}},
  \mnras, 440, 2442

\bibitem[{{Ricci} {et~al.}(2014{\natexlab{b}}){Ricci}, {Steiner}, \&
  {Menezes}}]{2014MNRAS.440.2419R}
{Ricci}, T.~V., {Steiner}, J.~E., \& {Menezes}, R.~B. 2014{\natexlab{b}},
  \mnras, 440, 2419

\bibitem[{{Richardson}(1972)}]{1972JOSA...62...55R}
{Richardson}, W.~H. 1972, Journal of the Optical Society of America
  (1917-1983), 62, 55

\bibitem[{{Schnorr M{\"u}ller} {et~al.}(2011){Schnorr M{\"u}ller},
  {Storchi-Bergmann}, {Riffel}, {Ferrari}, {Steiner}, {Axon}, \&
  {Robinson}}]{2011MNRAS.413..149S}
{Schnorr M{\"u}ller}, A., {Storchi-Bergmann}, T., {Riffel}, R.~A., {et~al.}
  2011, \mnras, 413, 149

\bibitem[{{Steiner} {et~al.}(2009){Steiner}, {Menezes}, {Ricci}, \&
  {Oliveira}}]{2009MNRAS.395...64S}
{Steiner}, J.~E., {Menezes}, R.~B., {Ricci}, T.~V., \& {Oliveira}, A.~S. 2009,
  \mnras, 395, 64

\bibitem[{{van Dokkum}(2001)}]{2001PASP..113.1420V}
{van Dokkum}, P.~G. 2001, \pasp, 113, 1420

\bibitem[{{Walcher} {et~al.}(2009){Walcher}, {Coelho}, {Gallazzi}, \&
  {Charlot}}]{2009MNRAS.398L..44W}
{Walcher}, C.~J., {Coelho}, P., {Gallazzi}, A., \& {Charlot}, S. 2009, \mnras,
  398, L44

\bibitem[{{Young} {et~al.}(2007){Young}, {Nowak}, {Markoff}, {Marshall}, \&
  {Canizares}}]{2007ApJ...669..830Y}
{Young}, A.~J., {Nowak}, M.~A., {Markoff}, S., {Marshall}, H.~L., \&
  {Canizares}, C.~R. 2007, \apj, 669, 830

\bibitem[{{Yusef-Zadeh} {et~al.}(2002){Yusef-Zadeh}, {Law}, {Wardle}, {Wang},
  {Fruscione}, {Lang}, \& {Cotera}}]{2002ApJ...570..665Y}
{Yusef-Zadeh}, F., {Law}, C., {Wardle}, M., {et~al.} 2002, \apj, 570, 665

\end{thebibliography}

\end{document}